\documentstyle[11pt,epsf]{article}

\begin{document}
\title{Quantum Mechanical Motion of Relativistic Particle in Non-Continuous Spacetime}
\author{Andreas Kull\thanks{266 East 78$^{th}$ Street, Apt.7, New York, NY 10021,
email:Andreas\_Kull@hotmail.com}}
\date{\today}
\maketitle

\begin{abstract}
The quantum mechanical motion of a relativistic particle in a
non-continuous spacetime is investigated. The spacetime model is a
dense, rationale subset of two-dimensional Minkowski spacetime.
Solutions of the Dirac equation are calculated using a generalized
version of Feynman's checkerboard model. They turn out to be
closely related to the continuum propagator.

\end{abstract}

\section{Introduction}

Continuity and a metric of Lorentz signature are fundamental
properties associated with spacetime in the framework of Special
and General Relativity. Though intuitive, the assumption of
continuity is debatable because of quantum mechanical arguments.
Discrete models of spacetime have been investigated thus by
various authors \cite{wen,das,mes,ya1,ya2}.

The modern idea that space could be discrete goes back to Riemann \cite{rie} who
reflected about a natural measure of space. More recent attempts to consider
spacetime as discrete either aim at resolving divergence problems in Quantum
Field Theory at a fundamental level or seek to reconcile General Relativity and
Quantum Theory to form a unified theory.

In this paper we investigate possible effects, which the
hypothetical discreteness of spacetime could have on Quantum
Theory. To this end, Feynman's path integral approach to Quantum
Tehory \cite{fey,fey2} provides an ideal framework since it
naturally accounts for spacetime properties. For continuous
spacetime, the path integral formalism has been shown to be
completely equivalent to ordinary wave mechanics. However, for a
non-continuous spacetime one should not necessarily expect
equivalent findings. To gain insight into the possible differences
between quantum mechanics in the framework of common continuous
spacetime and a non-continuous spacetime model, we seek solutions
for the Dirac equation.

As has been pointed out by Feynman and Hibbs \cite{fey}, the
retarded propagator of the $1+1$ dimensional Dirac equation
\begin{equation}
i{{\partial \Psi}/{\partial t}} = -i \sigma_{z} {{\partial
\Psi}/{\partial x}} - \sigma_{x} \Psi \label{diraceq}
\end{equation}
(with units $c=\hbar/m=1$)
can be obtained from a random walk model in which the particle
motion is restricted to movements either forward or backward at
the speed of light. We consider this model in the framework of a
particular non-continuous spacetime model \cite{ku1} and
demonstrate by an explicit calculation that the model yields
common solutions of the Dirac equation.

The paper is organized as follows: Section 2 gives a short review
of relevant properties of the non-continuous Minkowskain spacetime
model described in \cite{ku1}. In section 3 we calculate solutions
of the Dirac equation. Section 4 summarizes and discusses the
results.

\section{Non-Continuous Minkowskian Spacetime}
The two dimensional spacetime model considered here is defined by the subset $M$
of ${\bf R}^2$
\begin{eqnarray}
M = \{t,x\} &=&
\{ {{n}\over{m}}(p^2+q^2), {{n}\over{m}}(p^2-q^2) \} \subset {\bf R}^2,
\label{spacetime} \\
            & & n,m,p,q \in {\bf Z} \setminus\{0\}  \nonumber
\end{eqnarray}
In the following, $n,m,p,q$ may take any value of the indicated range and are
not to be considered as fixed. Elements of $M$ correspond to spacetime points
with temporal and spatial coordinates $(t,x)$. $M$ is rational,
hence countable, i.e. of cardinality $\aleph_0$ while ${\bf R}^2$  is of
cardinality $\aleph_1$. With respect to this difference we denote $M$ as
non-continuous.

\pagebreak
The set $M$ is invariant under the transformations
\begin{eqnarray}
\phi : M &\mapsto& M,  \\
\phi(s)&=&
{{1}\over{\sqrt{1-v^2}}}
\left( \begin{array}{cc}
 1 & -v \nonumber \\
 -v & 1
\end{array} \right) s  \label{phi} \nonumber \\
&=&
{{1}\over{2pq}}
\left( \begin{array}{cc}
 p^2+q^2 & -(p^2-q^2) \\
-(p^2-q^2) &  p^2+q^2
\end{array} \right) s ,   \nonumber  \\
& &p,q \in {\bf Z} \setminus\{0\}.   \nonumber
\end{eqnarray}
The parameter
\begin{equation}
\quad v=v(t,x)={{x}\over{t}}={{p^2-q^2}\over{p^2+q^2}}
\end{equation}
corresponds to the velocity defined as ratio of space $x$ and time $t$
displacements measured from the origin of the coordinate system. As illustrated
by the second line of (\ref{phi}), the transformation $\phi$ maps rational
spacetime coordinates $(t,x)$ onto rational ones $(t',x')$.

It is straightforward to verify that (i) the elements $\varphi \in
\phi$ map onto $M$, that (ii) that $\phi$ with respect to matrix
multiplications $\circ$ possesses group structure and that (iii)
$(\phi,\circ)$ is a subgroup of the common $1+1$ dimensional
Lorentz group (using natural units). According to these properties
and the non-continuity of $M$ we denote $M$ as {\it non-continuous
Minkowskian spacetime}. In \cite{ku1} it has been shown that in
the framework of the d-space formalism \cite{gru,he1,he2} the set
$M$ corresponds to a spacetime model which is nowhere
diffeomorphic to ${\bf R}^2$ but possesses the key properties of
the common macrophysical spacetime. In the same context $M$ may be
considered as a massless solution of the Einstein equations. Note
that in contrast to other discrete spacetime models (e.g.
\cite{ya1,ya2,hof} and references therein) which intrinsically
violate relativistic covariance and introduce an observationally
not supported minimal length, the spacetime model considered here
exhibits a generalized form of covariance. Since the set $M$ is
dense in ${\bf R}^2$, it doesn't introduce a minimal length as
well.

The following sections are based on the spacetime model
(\ref{spacetime}). We will consider light-cone coordinates
$(r,l)=({1\over{2}}(t+x),{1\over{2}}(t-x))$ in order to simplify
calculations. In light-cone coordinates $M$ becomes
\begin{eqnarray}
M^\ast = \{r,l\} = \{ {{n}\over{m}} {p^{2}}, {{n}\over{m}} {q^{2}} \}
\subset {\bf R}^2 \\
m,n,p,q \in {\bf Z} \setminus\{0\}.   \nonumber
\end{eqnarray}
To end this section we note that since the set $M$ is dense, it is
possible to formulate differential equations without referring to a calculus of
finite differences.

\section{Quadratic Checkerboard Model}
In \cite{fey} Feynman and Hibbs described a model for the quantum
mechanical motion of a relativistic electron. Solutions of the 1+1
dimensional Dirac equation are obtained by summing over all
possible particle trajectories consisting of movements either
forward or backward at the speed of light. Assuming natural units
$c=\hbar/m=1$, the motion of the particle corresponds to a
sequence of straight path segments of slope $\pm 45^\circ$ in the
x-t plane. The retarded propagator $\psi_{\delta\gamma}(x,t)$ of
the Dirac equation is obtained from the limiting process (e.g.
\cite{fey,jac})
\begin{equation}
\psi_{\delta\gamma}(x,t) = \lim_{N \rightarrow \infty}
A_{\delta\gamma}(\epsilon) \sum_{R \ge 0} N_{\delta\gamma}(R)
(\mbox{i} \epsilon)^{R} \label{proplin}
\end{equation}
$N$ is the number of segments with constant length $\epsilon = t/N$ between the
start point (which is assumed to be the origin of the coordinate system) and the
end point $(x,t)$ of the path. $R$ denotes the number of bends while
$N_{\delta\gamma}(R)$ stands for the total number of paths consisting of $N$
segments with $R$ bends. The indices $\gamma$ and $\delta$ correspond to the
directions forward or backward at the path's start and end points, respectively,
and refer to the components of $\psi$. $A_{\delta\gamma}(\epsilon)$ accounts for
the appropriate normalization.

Since the early 80'ies, the checkerboard model of relativistic
particle motion has been subject to anew interest (see e.g.
\cite{jac,ger,gav,or1,or2,mck,ku2}). In \cite{ku2} it has been
observed that paths with fixed start and end points have $R-1$
degrees of freedom, i.e. the last bend of a path is fully
specified by the location of the $R-1$ preceding bends and the end
point of the path. As well it has been shown that the
normalization constant $A_{\delta\gamma}(\epsilon)\equiv 1$ if
only the $R-1$ bends enter the calculation which actually define
the path of the particle. The general intuition behind this is the
idea that the origin of the divergence problems of Quantum Field
Theory could be related to an over-specification of the theory.
Here, the corresponding interpretation of the checkerboard model
and the calculation scheme described in \cite{ku2} are adapted.

In the following we demonstrate by an explicit calculation that
expression (\ref{proplin}) when generalized to account for the key
properties of $M$ yields solutions of the Dirac equation. Before
starting with the calculation, some remarks about the limiting
process (\ref{proplin}) and the terminology are in place. First
observe the structure of (\ref{proplin}).
$\psi_{\delta\gamma}(x,t)$ is defined in the limit $N \rightarrow
\infty$ where $N$ is the number of path segments a path has. The
length of each of these path segments is $\epsilon = t/N$. Bends
of the particle trajectory occur only at boundaries of path
segments. This is equivalent to consider a particle moving on a
rectangular spacetime lattice with equal spacing length (or
'resolution') $\epsilon$. Consider now the expression
\begin{equation}
\sum_{R \ge 0} N_{\delta\gamma}(R) (\mbox{i} \epsilon)^{R} \;.
\end{equation}
which is subject to the limit $N \rightarrow \infty$. For a given
$N$, it is weighted sum of the number $N_{\delta\gamma}(R)$ of
possible paths with $N$ segments and $R\ge0$ bends linking the
start and the end point under consideration. The complex weight
(amplitude) each path is contributing is $(\mbox{i}
\epsilon)^{R}$, i.e. each bend of a path contributes $(\mbox{i}
\epsilon)$. As has been demonstrated (e.g. \cite{fey,jac}), taking
the limit $N \rightarrow \infty$ yields solutions of the Dirac
equation.

\begin{figure}
\begin{center}
\leavevmode{\epsfxsize=10.5cm\epsffile{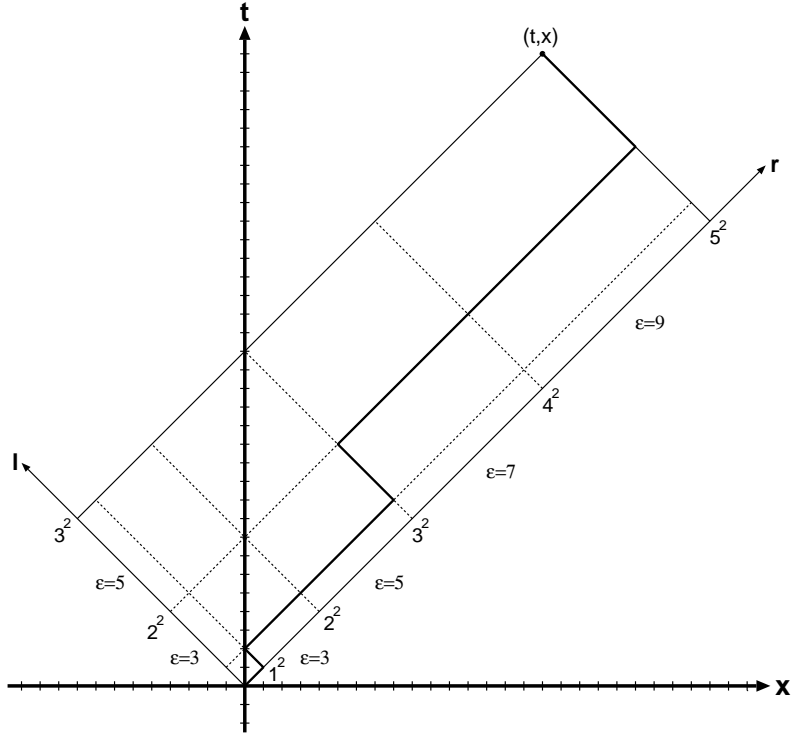}}
\end{center}
\caption{\label{fig1}A path on $M$ between the origin of the
coordinate system and a given end point $(t,x)$. The path has a
total of $N=P+Q=8$ segments, where $P=5$ segments are going to the
right and $Q=3$ segments to the left. A total of $R= R^+ + R^-=5$
bends occur, $R^+=2$ to the right and $R^-=3$ to the left. Note
that the last bend to the left is fully determined by the end
point $(t,x)$ and the location of $R-1=4$ bends, i.e. by the
$R^+=2$ bends to the right and the first $R^--1=2$ bends to the
left.}
\end{figure}

When adapting Feynman's checkerboard model to the discrete
spacetime $M$ we will follow as closely as possible the scheme
outlined above. Two differences have to be taken into account, the
first of which being a generalization. First, it is supposed that
path segments can be of non-uniform length $\epsilon_n \neq
\mbox{const.}$ and that the contribution of each bend to the
overall amplitude of a path is proportional to the segment length
$\epsilon_n$ immediately preceding the bend. This leads to the
generalized form
\begin{equation}
\psi_{\delta\gamma}(x,t) = \lim_{N \rightarrow \infty}
A_{\delta\gamma}(\epsilon) \sum_{R \ge 0} N_{\delta\gamma}(R)
(\mbox{i})^R \prod_{n=1}^R\epsilon_n \label{propquad}
\end{equation}
of (\ref{proplin}). Second, only particle paths on $M$ will be
considered. These path are characterized by bends occurring at
spacetime points $(t,x)$ satisfying
\begin{equation}
{x\over{t}}= {{p^2-q^2}\over{p^2+q^2}} \; .
\end{equation}
In terms of $M^\ast $, the light-cone representation of $M$,
this condition becomes
\begin{equation}
{r\over{l}}= {{p^2}\over{q^2}}
\label{condquad}
\end{equation}
where $(r,l)=({1\over{2}}(t+x),{1\over{2}}(t-x))$. Points of the
spacetime lattice should account for restriction (\ref{condquad}).
We achieved this by considering quadratic spacetime
lattices whose lattice points $(j_r,j_l)\in{\bf N}^2$ have, up to
scaling, spacetime lightcone coordinates $(r,l)=(j_r^2,j_l^2)$.
The spacetime coordinates of lattice points of this kind naturally
account for condition (\ref{condquad}). Figure \ref{fig1} shows an
example for a path with $N=8$ segments. With respect to the
modified properties of the spacetime lattice, the model is denoted
as {\it quadratic} checkerboard model.

From the definition of the quadratic spacetime lattice (or
directly from Figure \ref{fig1}) it is evident that path segments
$\epsilon_{n}$ are of non-uniform length and are depending on the
lattice location $(j_r,j_l)$. We only will be interested in the
length of paths segment preceding a bend and bends of a path to
the right and to the left will be considered separately in the
following. Observing that the length of a path segment preceding a
bend to the right (left) is uniquely specifying by the lattice
coordinate $j_r$ ($j_l$), the indexes $r,l$ can be dropped. Then,
the length of a path segment immediately following a bend can be
written as
\begin{equation}
\epsilon_{j}=(2 j-1)\epsilon_{0}
\label{quadepsilon1}
\end{equation}
The constant $\epsilon_0$ accounts for the scaling of the path. It
depends on the end point $(t,x)$ and on the number of segments $N$
a path has. For an explicit expression of $\epsilon_0$ in terms of
the number of segments to the left and right, see equation
(\ref{sclng}). Note that $\epsilon_0$ scales not just the length
of paths segment followed by a bend but the whole quadratic
spacetime lattice. From this it follows immediately that for $N
\rightarrow \infty$ the spacetime lattice under consideration
becomes a rectangular subset of $M$.

In line with the generalization (\ref{propquad}) of Feynman's
checkerboard model suppose now that each bend defining a path on
$M$ contributes an amplitude proportional to the length of the
path segment immediately preceding the bend. Again, bends to the
right and left are considered separately and the following
notation is adopted: The index $n$ enumerates the segments of a
path followed by a bend to the right (left). $j_n$ stands for the
lattice coordinate $j_r$ ($j_l$) of the segment and the set
$\{j_{n}\}$ indicates the path segments, after which bends to the
right (left) occur. For a schematic example consider Figure
\ref{fig1}. The length $\epsilon_{j_n}$ of the path segment
followed by the $n^{th}$ bend to the right (left) according to
(\ref{quadepsilon1}) becomes
\begin{equation}
\epsilon_{j_n}=(2 j_n-1)\epsilon_{0}
\label{quadepsilon2}
\end{equation}
and the corresponding contribution of the $n^{th}$ bend to the overall
amplitude is
\begin{equation}
\phi_{j_n}=\mbox{i} \epsilon_{j_n} = (2 j_n-1)\epsilon_{0}
\label{aquad1}
\end{equation}
The total amplitude of a path contributed by bends to the right
(left) is given by the product
\begin{equation}
\phi=\prod^{}_{n} (\mbox{i} \epsilon_{j_n})
\end{equation}

Having established this notation, we start with the evaluation of
(\ref{propquad}). As in the 'linear' case (\cite{ku2}) a path with
$R$ bends that starts with a positive velocity (i.e. to the right)
and ends with a negative velocity (i.e. to the left) consists of
exactly $(R-1)/2+1$ bends to the left and $(R-1)/2$ bends to the
right. The $(R-1)/2$ bends to the right can occur after an
arbitrary path segment to the left. $(R-1)/2$ of the $(R-1)/2+1$
bends to the left occur in the same manner after path segments to
the right while the additional bend to the left must occur after
the last segment to the right. Let $P$ denote the total number of
path segments to the right $(+)$ and $Q$ those to the left $(-)$.
In total, the path has $N=(P+Q)$ segments. The contribution of the
$R^+=(R-1)/2$ bends to the right $\psi_{-+}$ is
\begin{equation}
\psi_{-+}(R^+)
        = \sum^{P-1}_{j_1 < \ldots < j_{R^+}} (2 j_1-1)\cdot\ldots\cdot(2
j_n-1)\cdot(\mbox{i}\epsilon_{0})^{R^+}
\end{equation}
Next consider the situation where the path consists of a large
number of segments to the right, i.e. $P \gg 1$ or equivalently $N
\rightarrow \infty$. This limiting process corresponds to the
spacetime lattice becoming dense. For $P \gg 1$, $\psi_{-+}(R^+)$
is approximated by
\begin{eqnarray}
\psi_{-+}(R^+) &\approx& {1\over{{R^+}!}}\sum^{P}_{j_1 \neq\ldots\neq j_{R^+}}
(j_1 \cdot\ldots\cdot j_n)\cdot 2^{R^+}(\mbox{i}\epsilon_{0})^{R^+} \\
        &\approx& { 2^{R^+} (\mbox{i}
\epsilon_{0})^{R^+}\over{{R^+}!}}\left( \sum_{j=1}^{P} j \right) ^{R^+} \\
        &\approx& { 2^{R^+} (\mbox{i}
\epsilon_{0})^{R^+}\over{{R^+}!}}\left( {P^2\over{2}}\right) ^{R^+} \\
        &=& { P^{2R^+} (\mbox{i}\epsilon_{0})^{R^+}\over{{R^+}!}}
\end{eqnarray}
The contribution of the $R^-=(R-1)/2+1$ bends to the left is
calculated similarly. The additional bend (occurring after the
last segment to the right) does not enter the calculation since a
path is fully determined by the $R-1$ bends to the right and left,
respectively. We find
\begin{eqnarray}
\psi_{-+}(R^-) &\approx& { 2^{(R^- -1)} (\mbox{i}
\epsilon_{0})^{(R^- -
1)}\over{{(R^- -1)}!}}\left( {Q^2\over{2}}\right) ^{(R^- -1)}\\
        &=& { Q^{2(R^- -1)} (\mbox{i} \epsilon_{0})^{(R^- -1)}\over{{(R^- -1)}!}}
\end{eqnarray}
To get the total contribution of the $R^-$ bends to the left and
the $R^+$ bends to the right, expressions $\psi_{-+}(R^-)$ and
$\psi_{-+}(R^-)$ are multiplied yielding
\begin{equation}
\psi_{-+}(R)=(\mbox{i}
\epsilon_0)^{R-1}{(PQ)^{R-1}\over{[((R-1)/2)!]^2}}
\label{rightleft}
\end{equation}
which represents the contribution of all paths with exactly $R$
bends that start to the right and end to the left. Finally, the
summation over all possible paths that start to the right and end
to the left is preformed by summing $\psi_{-+}(R)$ over the total
number of bends $R$. This leads to
\begin{equation}
\psi_{-+}=\sum_{\mbox{\tiny odd } R} (\mbox{i} \epsilon_0)^{R-1}
{(PQ)^{R-1}\over{[((R-1)/2)!]^2}} \label{psimp}
\end{equation}
where
\begin{equation}
\epsilon_0={t\over{P^2+Q^2}} \label{sclng}
\end{equation}
accounts for the proper scaling of the lattice. Notice that for $N
\rightarrow \infty$ the scaling factor $\epsilon_0 \rightarrow 0$.
Defining the classical velocity of the particle $v={\Delta
x/{\Delta t}}=x/t={{(P^2-Q^2)}/{(P^2+Q^2)}}$ we find
$PQ=(P^2+Q^2)/2\gamma$ where $\gamma=1/\sqrt{1-v^2}$. Finally
$\psi_{-+}$ becomes
\begin{eqnarray}
\psi_{-+}
&=& \sum^{\infty}_{k=0}(-1)^k {({t /{2\gamma}})^{2k}\over{[(k)!]^2}} \\
&=& J_0(t/\gamma) \label{psipm}
\end{eqnarray}

A similar calculation of $\psi_{+-}$ leads to the same result.
This can be seen from interchanging the roles of $P$, $Q$, $R^-$
and $R^+$.

For $\psi_{++}$, the number of bends to the right and to the left is $R/2$ for
each direction where $R$ is even. However, the path again is defined by
$R^+=R/2$ bends to the right and $R^-=R/2-1$ to the left. Thus,
\begin{eqnarray}
\psi_{++} &=& \sum_{\mbox{\tiny even } R} (\mbox{i}
\epsilon_0)^{R-1} {P^R Q^{R-2}\over{(R/2)!(R/2-1)!}} \\
&=& \sum_{0,2,4,\ldots} (\mbox{i} \epsilon_0)^{R+1} {P^2
(PQ)^R\over{(R/2+1)!(R/2)!}} \\
&=& {P\over{Q}}\sum^{\infty}_{k=0} (-1)^k {(PQ
\epsilon_0)^{2k+1}\over{(k+1)!(k)!}} \\
&=& \mbox{i}{P\over{Q}}\sum^{\infty}_{k=0} (-1)^k {(
t/2\gamma)^{2k+1}\over{(k+1)!(k)!}} \\
&=& \mbox{i}{P\over{Q}} J_1(t/\gamma) \; .
\end{eqnarray}
With $P/Q = (x+t)/(t^2-x^2)^{1/2}$ the component $\psi_{++}$ becomes
\begin{equation}
\psi_{++}= \mbox{i}{(t+x)\over{t}}\gamma J_1(t/\gamma) \, .
\label{psipp}
\end{equation}
A similar calculation for $\psi_{--}$ leads to
\begin{equation}
\psi_{--}= \mbox{i}{(t-x)\over{t}}\gamma J_1(t/\gamma) \, .
\label{psimm}
\end{equation}
which completes the calculation.

To relate the components $\psi_{\delta\gamma}$ to the Dirac
equation (\ref{diraceq}) consider its explicit representation with
\begin{equation}
\sigma_x=
\left( \begin{array}{cc}
   0 &  1 \\
   1 &  0
\end{array} \right), \quad
\sigma_z =
\left( \begin{array}{cc}
   1 &  0 \\
   0 & -1
\end{array} \right) \; .
\end{equation}
In this representation, $\psi_1$ and  $\psi_2$ defined as
\begin{equation}
\psi_1=
\left( \begin{array}{c}
   \psi_{++} \\
   \psi_{+-}
\end{array} \right), \quad
\psi_2 =
\left( \begin{array}{c}
   \psi_{+-} \\
   \psi_{--}
\end{array} \right)
\end{equation}
are two common independent solutions of the Dirac equation
(\ref{diraceq}). Since $M$ is dense in ${\bf R}^2$, the
differential form of (\ref{diraceq}) does not pose a problem even
though $M$ is non-continuous.

The four components $\psi_{\delta\gamma}$ are closely related to
the matrix elements of the retarded continuum propagator of the
Dirac equation (see e.g. \cite{jac}). Despite this there is no
full equivalence and a physical interpretation of
$\psi_{\delta\gamma}$ in terms of propagator components is not
straightforward. In the last section we will come back to this and
related questions in more detail.

As a side remark note that for the applied calculation scheme it
is essential to consider only those bends which actually define
the path of the particle. While it is straightforward to
'renormalize' Feynman's original model if all bends are
considered, this is not the case here. Taking into account all
bends, the normalization `constant'
$A_{\delta\gamma}(\epsilon_{0})$ for the checkerboard restricted
to $M$ turns out to not only depend on $\epsilon_{0}$ but also on
$\delta$, $\gamma$ and the classical velocity $v$.

\section{Discussion and Conclusions}
It is counter-intuitive that Feynman's spacetime based path
integral approach to quantum theory yields in the case of the
discussed example intriguing similarities between the common
continuous and the discrete spacetime model $M$. The solutions
$\psi$ obtained from the generalized checkerboard model and the
common retarded continuum propagator of the Dirac equation are of
the same form.

However, despite the same functional form, the components
$\psi_{\delta\gamma}$ are not fully equivalent to the retarded
continuum propagator. Reworking the outlined calculation with
physical units $\hbar/m\not=1$ shows the dimensionality of
$\psi_{\delta\gamma}$ to be $lenght^{0}$ which is different from
the dimensionality $lenght^{-1}$ of the retarded continuum
propagator. With respect to the original checkerboard model, the
difference is related to two origins: First, taking into account
only $R-1$ bends means accounting for one factor $m\epsilon$ less.
Second, Feynman's original checkerboard model is 'renormalized' by
dividing components like e.g. (\ref{psimp}) and (\ref{psipp}) by
$2 \epsilon$ in order to yield the retarded continuum propagator.

The dimensionality $lenght^{0}$ of $\psi_{\delta\gamma}$ may be
seen as suggesting a picture consistent with the cardinality
$\aleph_0$ of $M$. Profoundly taking into account the cardinality
$\aleph_0$, a continuum-like propagator is inconsistent with $M$.
This is due to the fact that because of its cardinality there
exists no measure on $M$. As a consequence, the concept of
integration is not available. Instead, one can expect (infinite)
sums to play the role of integration. This suggests that the
components of a discrete counterpart of the continuum propagator
(i.e. $\psi_{\delta\gamma}$) should have dimensionality
$lenght^{0}$. Fully clarifying the physical meaning of
$\psi_{\delta\gamma}$ in the framework of the discrete spacetime
model $M$ is not straightforward. One is lead into number
theoretical problems that are not easy to tackle.

Other open questions remain. For example: To what extend does the
path integral approach capture the relation between spacetime and
quantum mechanics? This question is raised not only by the present
results but already by Feynman's original checkerboard model.
After all, according to its formulation, bends occur only at
spacetime points with rational coordinates. As well, the velocity
spectrum of the particle is rational. In this sense, Feynman's
checkerboard model is of cardinality $\aleph_0$ and thus discrete
in the same manner as $M$ is. To our knowledge this is a fact that
the literature has not paid attention to.

In \cite{kau} it has been shown that the solution of the Dirac
equation can be understood in terms of bit strings detached from
any notion of spacetime. The present results seem to point into
the same direction, namely, that there is (at least in the special
case of the Dirac equation) only a loose connection between
spacetime properties and quantum mechanics, except probably for
the signature of the metric.

Other open issues are related to the spacetime model $M$ itself.
The ratio $x/t$ interpreted as classical velocity of the particle
(starting from the origin) implies a discrete velocity spectrum
\begin{equation}
\{v\}={{p^2-q^2}\over{p^2+q^2}},\quad p,q \in {\bf Z} \setminus\{0\}.
\label{vspect}
\end{equation}
What are the implications of the non-continuous spectrum
(\ref{vspect})? Progress in this and other directions again is
complicated by the fact that $M$ is of cardinality $\aleph_0$.
Finally, a question not addressed here is where the bends of a
path occur, that contribute most to the overall amplitude. Because
of the structure of $M$ it is not evident that the locations are
distributed as in the continuous case.

In summary, it has been shown that Feynman's path integral
approach restricted to the subset $M$ of ${\bf R}^2$ recovers
common solutions of the Dirac equation. The particle path on $M$
is limited to bends at positions $i$ with coordinate ratios
$x_{i}/t_{i}={({p^2-q^2})/({p^2+q^2})}$ where $p,q \in {\bf Z}
\setminus\{0\}$. The bends of paths on $M$ thus occur at a subset
of the rational space and time coordinates only. It should be
noted that both the characteristics of $M$ and the discrete
velocity spectrum (\ref{vspect}) are invariant under the Lorentz
subgroup $\phi$. In particular, bends of paths occur only at
rational space and time coordinates and the velocity remains
rational in all coordinate systems. In this sense the model
exhibits a generalized form of Lorentz invariance missing in
Feynman's original checkerboard model and most other lattice
models of spacetime. This demonstrates that there is a discrete,
two-dimensional spacetime model of cardinality $\aleph_0$
accounting for some of the key properties of special relativity
and quantum mechanical features of relativistic particle motion.

\section*{Acknowledgements}
The author thanks an unknown referee for helpful comments.

\pagebreak

\end{document}